\documentclass[10pt,prb,aps,twocolumn,showpacs,floatfix,superscriptaddress]{revtex4-1}
\usepackage{graphicx}
\usepackage{amssymb}
\usepackage{amsmath}
\usepackage{color}
\usepackage[caption=false]{subfig}
\newcommand{\bc}{\begin{center}}
\newcommand{\ec}{\end{center}}
\newcommand{\be}{\begin{equation}}
\newcommand{\ee}{\end{equation}}
\newcommand{\ba}{\begin{array}}
\newcommand{\ea}{\end{array}}
\newcommand{\beq}{\begin{eqnarray}}
\newcommand{\eeq}{\end{eqnarray}}
\newcommand{\ket}[1]{\left| {#1}\right\rangle}
\newcommand{\bra}[1]{\left\langle {#1} \right|}
\newcommand{\skp}[2]{\langle {#1} | {#2}\rangle}
\newcommand{\e}[1]{\langle {#1}\rangle}
\newcommand{\abs}[1]{\left| {#1}\right|}
\newcommand{\trip}[3]{\langle {#1}|{#2}|{#3}\rangle}

\begin{document}

\title{Properties of the random-singlet phase: \\
from the disordered Heisenberg chain to an amorphous valence-bond solid}

\author{Yu-Rong Shu}
\affiliation{State Key Laboratory of Optoelectronic Materials and Technologies,
School of Physics, Sun Yat-Sen University, Guangzhou, China}

\author{Dao-Xin Yao}
\email{yaodaox@mail.sysu.edu.cn}
\affiliation{State Key Laboratory of Optoelectronic Materials and Technologies,
School of Physics, Sun Yat-Sen University, Guangzhou, China}

\author{Chih-Wei Ke}
\affiliation{Graduate Institute of Applied Physics, National Chengchi University, Taipei, Taiwan}

\author{Yu-Cheng Lin}
\email{yc.lin@nccu.edu.tw}
\affiliation{Graduate Institute of Applied Physics, National Chengchi University, Taipei, Taiwan}

\author{Anders W. Sandvik}
\email{sandvik@bu.edu}
\affiliation{Department of Physics, Boston University, 590 Commonwealth Avenue, Boston, Massachusetts 02215, USA}

\begin{abstract}
We use a strong-disorder renormalization group (SDRG) method and ground-state  quantum Monte Carlo (QMC) 
simulations to study $S=1/2$ spin chains with random couplings, calculating disorder-averaged spin and dimer 
correlations. The QMC simulations demonstrate logarithmic corrections to the power-law decaying correlations 
obtained with the SDRG scheme. The same asymptotic forms apply both for systems with standard Heisenberg 
exchange and for certain multi-spin couplings leading to spontaneous dimerization in the clean system. We
show that the logarithmic corrections arise in the valence-bond (singlet pair) basis from a contribution
that can not be generated by the SDRG scheme. In the model with multi-spin couplings, where the clean system
dimerizes spontaneously, random singlets form between spinons localized at domain walls in the presence of disorder. 
This amorphous valence-bond solid is asymptotically a random-singlet state and only differs from the random-exchange 
Heisenberg chain in its short-distance properties.
\end{abstract}

\date{\today}

\maketitle

\section{Introduction}
\label{sec:intro}

A remarkably simple but powerful method was introduced some time ago by Ma {\it et al.}~for studies of quantum magnets with random 
couplings:\cite{ma79} In a repeated decimation procedure that gradually lowers the energy scale, 
the strongest coupled spin pair is identified and put into a singlet state, 
which decouples from the rest of the system after new effective couplings are generated among the remaining spins. This 
{\it strong-disorder renormalization group} (SDRG) scheme often flows toward a random singlet (RS) fixed-point,\cite{fisher92} which is 
universal for a broad class of spin chains.\cite{fisher94} The SDRG method has become a standard tool for studying a wide range of systems 
\cite{bhatt82,westerberg97,fisher98,motrunich00,hikihara99,melin02,refael02,refael04,igloi05,schehr06,hoyos08,altman10,iyer12,pielawa13,monthus15} 
and the RS phase represents a corner-stone of our understanding of disorder in quantum many-body physics. 

\begin{figure}
\centerline{\includegraphics[width=8cm, clip]{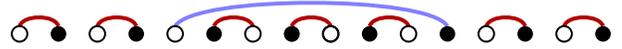}}
\vskip-2mm
\caption{(Color online) Qualitative AVBS ground state of an $S=1/2$ spin chain. The open and solid circles represent the two sublattices of 
the bipartite lattice and the arches indicate singlets (valence bonds). The short valence bonds form ordered domains, between
which spinons localize. In the ground state the spinons freeze pairwise into long-bond singlets.}
\label{fig1}
\vskip-3mm
\end{figure}

Here we compare SDRG calculations and ground-state projector quantum Monte Carlo (QMC) simulations in the valence-bond (VB) basis
for two types of $S=1/2$ spin chains which in the absence of disorder have very different ground states; the standard quasi-ordered 
(critical) Heisenberg antiferromagnet with nearest-neighbor exchange and a chain with multi-spin interactions that lead to a spontaneously 
dimerized (VB solid, VBS) ground state. In the latter case, in the presence of disorder, we demonstrate an amorphous VBS (AVBS) with out-of-phase 
dimerized chain segments separated by spin-carrying domain walls, as illustrated in Fig.~\ref{fig1}. Despite the very different local
properties of the two systems, they both exhibit characteristic RS properties in their long-distance correlations. 

%%%%%%%%%%%%%%%%% FIG1 %%%%%%%%%%%%%%%%%%%%%%%
%%%%%%%%%%%%%%%%%%%%%%%%%%%%%%%%%

In addition to the spin-spin correlation function, on which past analytical and numerical calculations have been
focused, we also compute the dimer-dimer correlation functions (where the dimer operator measures the density of
singlets on a nearest-neighbor bond). The SDRG spin-spin correlations are known to decay asymptotically with 
distance $r$ as $r^{-2}$, and this behavior is very well reproduced by numerically iterating the SDRG procedures.
We here find numerically that the SDRG dimer-dimer correlations decay as $r^{-4}$. Surprisingly, in light of the large
number of previous studies and the widely accepted notion that the $r^{-2}$ asymptotic form for the mean spin-spin correlations is
exact,\cite{fisher94} our QMC results show that the SDRG method misses universal multiplicative logarithmic (log) 
corrections to the power-law decays. Similar corrections were noted \cite{igloi00,ristivojevic12} in other classes of disordered 
spin chains but had not been anticipated in the present case. By studying different contributions to the spin-spin correlation functions 
in QMC calculations in the VB basis, we find that the log corrections arise from a contribution that is completely missing in the simple 
singlet-product ground state resulting from the SDRG method. We find the same multiplicative logs both in the standard random-$J$ model 
and the random-$Q$ model (i.e., in the AVBS), thus reinforcing our claim that these corrections constitute a universal characteristic 
of the RS phase for SU(2) spin chains.

The outline of the rest of the paper is as follows: In Sec.~\ref{sec:models} we define the models and outline the SDRG and QMC methods.
We also show results for the energy flow in the SDRG calculations for both the random $J$ and random $Q$ models, demonstrating the same asymptotic
flow in both cases, but with interesting cross-overs between different decimation stages for the random-$Q$ model. In Sec.~\ref{sec:col} we compare
SDRG and QMC results for spin-spin and dimer-dimer correlations in the random-$J$ model and discuss the origin of the log corrections in the VB basis. 
The random-$Q$ model and its AVBS state are discussed in Sec.~\ref{sec:avbs}. We briefly summarize our conclusions and provide some further remarks on the
significance of our findings in Sec.~\ref{sec:discussion}. Technical details of the SDRG scheme in the presence of the $Q$ term are presented in 
Appendix \ref{app:sdrg}, and in Appendix \ref{app:comp} we tabulate numerical resuls for the spin-spin correlations and discuss minor discrepancies 
with previous calculations.

\section{Models and methods}
\label{sec:models}

We consider interactions written with singlet projectors on two spins $i,j$,
\be
P_{i,j}=1/4-{\bf S}_i \cdot {\bf S}_j.
\ee
The antiferromagnetic Heisenberg Hamiltonian for a chain with $N$ spins can be written as
\be 
  H_J=-\sum_{i=1}^N J_iP_{i,i+1},
  \label{eq:J}
\ee 
where $J_i>0$ is a random antiferromagnetic
coupling. To achieve a robust VBS state (an AVBS in the presence of disorder) accessible to QMC calculations 
without sign problems, we use the six-spin interaction \cite{tang11a} to construct a chain described by 
\be
  H_Q=-\sum_{i=1}^N Q_i P_{i,i+1}P_{i+2,i+3}P_{i+4,i+5}
  \label{eq:Q}
\ee
%summed over all $i$ 
with random $Q_i>0$. A similar four-spin coupling also leads to a VBS, but with a
much smaller order parameter. Note that the clean Hamiltonian ($Q_i =1~ \forall i$) is 
translationally invariant and the system dimerizes by spontaneous symmetry-breaking, leading to a 
two-fold degenerate ground state. In both models we use periodic boundary conditions and the following 
distribution of the random couplings ($\lambda=J_i$ or $\lambda=Q_i$):
\begin{equation}
\pi(\lambda) = \left\lbrace 
\begin{array}{ll}d^{-1}\lambda^{1/d-1}, & {\rm ~for~} 0 < \lambda \le 1, \\ 0, & {\rm ~else,} \end{array}\right.
\label{dist}
\end{equation}
which is uniform within the range $(0,1]$ for $d=1$ and becomes singular when $d\to \infty$. 

\subsection{Strong-disorder RG}

The basic idea of the strong-disorder renormalization-group (SDRG) scheme is to find a system's ground state by successively eliminating
degrees of freedom with high energy. The SDRG method for the random Heisenberg chain (the random-$J$ model) is well documented and we refer 
to the literature for details.\cite{ma79,fisher92,fisher94,bhatt82,igloi05} In essence, the RG procedure for the random Heisenberg chain
consists of iteratively locating the 
two spins connected by the strongest coupling $\Omega=\text{max}\{J_i \}$, putting these in their singlet ground state and perturbatively generating 
an effective coupling between the neighboring spins with strength
\be
      \tilde{J}=\frac{J' J''}{2\Omega} < \Omega,\, J',\, J'', 
      \label{eq:XXX}
\ee
where $J'$ and $J''$ are couplings between the singlet and the neighboring spins. One can also do this step non-perturbatively by diagonalizing
the relevant subspace exactly,\cite{hikihara99} but this does not change the asymptotic behavior. The decimated spins are now ``frozen out'' and
will form a VB in the ground state that is successively generated by repeating the steps. This process yields an effective Hamiltonian 
with gradually fewer degrees of freedom and lower energy scale. For the antiferromagnets considered here, the final ground state is a product of 
singlet pairs, i.e., a single VB configuration.

The generalization of the SDRG to the random-$Q$ model (\ref{eq:Q}) with three singlet projectors (also called $Q_3$ interactions) is non-trivial, 
as the multi-spin interaction generates various terms of the forms $P_{i,i+1}P_{i+2,i+3}$ ($Q_2$ interactions) and $P_{i,j}$ ($J$ interactions) under 
SDRG, with several different cases in the perturbative treatment of the decimated operators. The technical details of the method is described in 
Appendix~\ref{app:sdrg}. Here we comment on the energy flows and demonstrate identical asymptotic behaviors for the random-$J$ and random-$Q$ systems.

Since the decimation procedure applied in the SDRG is an approximation relying on the flow toward a singular coupling distribution, the method is 
in general not suitable for studying systems where the quenched disorder is irrelevant in the renormalization-group sense. For systems governed by 
strong disorder, the approximation made in perturbation calculations and the ``freezing'' of degrees of freedom becomes inconsequential in the 
long-distance limit; in essence, because these systems, when studied at ever larger length scales (lower energy), appear more and more 
disordered. The RS state, which is the SDRG solution for the approximate ground state of the random Heisenberg chain (as well as many other spin
chains, e.g., the random XX-chain \cite{fisher94}), is a prominent example for extremely strong randomness, called infinite randomness fixed-point 
solutions. The fixed point is characterized by unconventional dynamic scaling, 
\be
 \ln \xi_t \sim \xi^\psi,
 \label{eq:exp_sc}
\ee
of the correlation length $\xi$ and the correlation time $\xi_t$, implying an infinite dynamic exponent, in contrast to the conventional power-law 
scaling,
\be
  \xi_t\sim\xi^z,
 \label{eq:pow_sc}
\ee
with a finite dynamic exponent $z<\infty$. In SDRG, the dynamic scaling behavior can be identified, for example, by examining the RG flow of the 
logarithmic energy scale $\ln(\Omega)$; here the energy scale is the strongest effective coupling at each step
of RG.

%%%%%%%%%%%%%%%%%%%%%%%%%%%%%%% FIG2  %%%%%%%%%%%%%%%%%%%%%%%%%%%%%%%%%
\begin{figure}[t]
\includegraphics[width=8.3cm, clip]{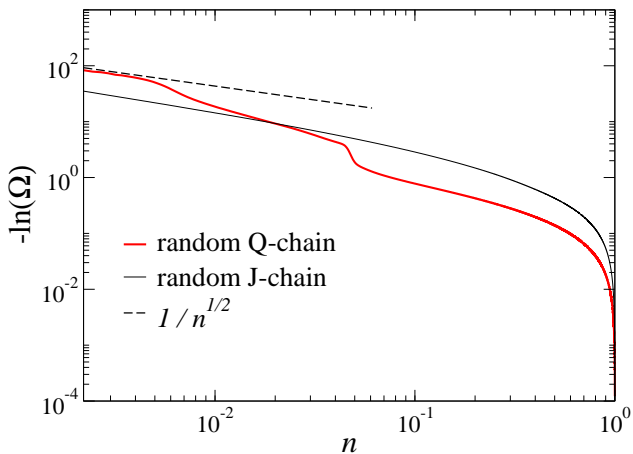}
\vskip-2mm
\caption{\label{fig:SDRG_gaps}
(Color online) SDRG evolution of the disorder-averaged log-energy scale for the random-$Q$ chain (thicker red curve) and the random-$J$ 
chain (thinner black curve), both with disorder parameter $d=1$ and chain length $N=8192$. The energy is graphed versus the fraction 
$n=N_\Omega/N$ of the active (not yet decimated) spins under the action of the RG (thus, $n=2/N$ at the final step). Asymptotically the curves for 
the both models tend to a power-law $-\ln(\Omega)\sim 1/\sqrt{n}$ in the late stage of the RG, as indicted by the dashed line.}
\vskip-2mm
\end{figure}

%%%%%%%%%%%%%%%%%%%%%%%%%%%%%%%%%%%%%%%%%%%%%%%%%%%%%%%%%%%%%%%%%%%%%%%

Fig.~\ref{fig:SDRG_gaps} shows the RG evolution of the log-energy scales for the two systems considered in this work, both with chain 
length $N=8192$ and disorder parameter $d=1$. In both these cases, the log-energy scale tends to a power-law relation with the number ${N}_\Omega$ 
of the active spins (the spins that are not yet decimated at a given energy scale $\Omega$) as
\be
   -\ln(\Omega) \sim {N}_\Omega^{-1/2},
   \label{eq:omega_N}
\ee
corresponding to a non-power-law dynamic scaling given in Eq.~(\ref{eq:exp_sc}) with $\psi=1/2$ and an infinite dynamic exponent $z\to \infty$, 
as predicted for an RS state.\cite{fisher94}

The convergence of the energy scale for the random $Q$-chain in Fig.~\ref{fig:SDRG_gaps} is slower than for the random-$J$ chain. Furthermore, 
the evolution for the random $Q$-chain exhibits an interesting three-stage structure, which correspond to predominant $Q_3$-decimation in the early 
stage, mixed $Q_2$- and $J$-decimation in the intermediate stage, and predominantly $J$-decimation in the late stage. The ultimately same asymptotic
energy flows already is an indication of both systems flowing to the same RS fixed point. In later sections we will present QMC calculations
demonstrating this in an unbiased (non-approximate) way using correlation functions.

In comparison to the RS phase in the random XX chain,\cite{igloi05} the energy-length relation for the random Heisenberg chain shows slower convergence 
to the fixed-point solution given in Eq.~(\ref{eq:omega_N}) due to the factor $1/2$ in the recursion relation Eq.~(\ref{eq:XXX}), which is missing in 
the corresponding recursion relation for the XX-chain.\cite{fisher94} For the random $Q$-chain studied here, the recursion relations are more complex 
and factors like $1/32$ appear in the effective couplings (see Appendix~\ref{app:sdrg}); therefore the convergence is even slower than for the 
random-$J$ case.

\subsection{Projector QMC}
\label{sec:QMC}

For the QMC calculations, we employ a ground-state projection technique operating in the VB
basis.\cite{sandvik05,sandvik10} For our unfrustrated systems with bipartite interactions, we choose
a restricted VB basis in which all bonds connect sites on different sublattices; we denote such a basis vector 
by  
\be
   \ket{v}=\bigotimes_{i\in\mathcal{A},j\in\mathcal{B}} \ket{(i, j)},
\ee
where
\be
    \ket{(i,j)}=\frac{1}{\sqrt{2}}\left(\ket{\uparrow_i \downarrow_j}-\ket{\downarrow_i \uparrow_j} \right)
\ee
is the singlet state of two spins $i,\,j$ in different sublattices $\mathcal{A}$ and $\mathcal{B}$.
This basis is ideal for singlet ground states, as $S>0$ excitations 
are excluded from the outset, unlike finite-temperature methods which include the full Hilbert space. 
In the present case, the convergence to the ground state is accelerated by projecting from ``trial 
states'' obtained using the SDRG method for each set of random couplings. 

For the projector filtering out the ground state from the trial state, we use a 
power $(-H)^m$ of the Hamiltonian and carefully check for convergence as a function of $m$. Individual
strings of operators contributing to $(-H)^m$ are sampled, with each such string of terms in $H_J$ and $H_Q$ 
forming a long list  (of between $m$ and $3m$ elements) of singlet projectors $P_{i,j}$, each successively 
acting on two spins in a VB state and propagating this state according to
\be
 \begin{split}
    P_{i,j}\ket{\cdots (i,j) \cdots} &=\ket{\cdots (i,j) \cdots}, \\
    P_{i,j}\ket{\cdots (i',i) (j,j') \cdots}&= \frac{1}{2}\ket{\cdots (i,j) (i',j')\cdots}.
  \end{split}
  \label{eq:P_action}
\ee
In practice, the VB basis is explicitly used only when collecting measurements of the quantities computed. Here an advantage 
of the VB basis is the easy access to correlation functions expressed in terms of transition-graph loops.\cite{liang88,beach06,tang11b}
The sampling of the operator strings is most efficiently done by re-expressing the operator strings and the VB trial state are re-expressed
in the conventional basis of spin-$z$ components, where a powerful loop algorithm can be employed.\cite{sandvik10}

\section{Correlations in the RS phase}
\label{sec:col}

Field theory approaches predict multiplicative logarithmic corrections to a power law decay of correlation functions in the clean Heisenberg antiferromagnetic chain at zero temperature. The staggered spin-spin correlation function behaves as \cite{affleck89,singh89,giamarchi89,sandvik93} 
\be
     C(r)=(-1)^r\e{{\bf S}_i \cdot {\bf S}_{i+r}} \sim \frac{\ln^{1/2}(r)}{r},
     \label{eq:C_clean}
\ee
where logarithmic correction appears due to a marginally irrelevant operator in the field theory description.\cite{affleck89,singh89,giamarchi89}
Also the dimer correlation function, defined as 
\be
   D(r)=\langle B_i \cdot B_{i+r}\rangle - \langle B_i\rangle^2,
\label{eq:D_def}
\ee
where $B_i={\bf S}_i \cdot {\bf S}_{i+1}$, acquires a log correction and decays with distance as \cite{giamarchi89}
\be
     D(r)\sim (-1)^r\frac{ln^{-3/2}(r)}{r},
\ee
i.e., with only the power of the log correction being different from that in the spin correlations.

One of the key analytical SDRG results in one dimension
is that the long-distance staggered {\it mean} spin-spin correlation function $C(r)$
decays with distance $r$ as 
\be 
   C(r)\sim \frac{1}{r^{2}}
   \label{eq:C_SDRG} 
\ee
in the RS phase,\cite{fisher94} in contrast to Eq.~(\ref{eq:C_clean}) for the clean Heisenberg chain, Eq.~(\ref{eq:C_SDRG}) is obtained by 
assuming that the mean spin correlation function is dominated by rare long VBs in the ground state, a consequence of the distribution of VB lengths 
$P(\ell)$ at $\Omega\to 0$ in the SDRG framework, which to leading order has an inverse-square form as a function of the bond length 
$\ell$:\cite{fisher94,hoyos07}
\be
   P(\ell) \sim \frac{1}{\ell^{2}}.
   \label{eq:P_length}
\ee
Different from the rare events, a typical pair of widely separated spins, $i$ and $j$, do not form a singlet and the correlation between two such 
spins decays exponentially with the distance:
\be
   C^{\text{typ}}(\abs{i-j}) \sim \exp(-c \sqrt{\abs{i-j}}).
   \label{eq:C_typ}
\ee 
This behavior can be computed within the SDRG procedures by perturbatively taking into account the neglected correlations mediated 
by the decimated singlets.\cite{fisher94}

The asymptotic $r^{-2}$ decay of the mean spin-spin correlation in the RS phase has been tested by numerical calculations in spin chains with anisotropic 
interactions, in particular in the random XX chain,\cite{henelius98,hoyos07} which can be mapped to free fermions. Systems with isotropic Heisenberg interactions 
have proved much more challenging and the available numerical evidence is less convincing.\cite{hoyos07,laflorencie04,hikihara99,carlon04,goldsborough14}
In previous works it was implicitly assumed that the asymptotic form should be exactly $\propto 1/r^2$, as in the SDRG. Logarithmic corrections have been 
predicted and found in some types of random quantum spin chains,\cite{igloi00,ristivojevic12} but so far no such corrections have been considered in 
the case of random Heisenberg chain. Here we reach larger system sizes than in previous QMC studies and we also impose strict convergence controls, to
ensure that true ground state properties are obtained. The results reach a level of precision where we can unambiguously detect deviations
from the expected behavior that cannot be explained by standard higher-order power-law corrections. It is then natural to consider log corrections,
and, indeed, we find strong evidence for their presence in QMC results for $d=1$ and $d=2$ in the coupling distribution (\ref{dist}).

We first use the unbiased zero-temperature QMC method described in Sec.~\ref{sec:QMC} to investigate the spin correlation in the RS phase of the 
Heisenberg chain. We detect the multiplicative log-correction to the universal inverse-square law by comparing the data with the correlation 
obtained by the numerical SDRG. In addition, we have computed the dimer-dimer correlation function, which to our knowledge has not been previously 
considered, neither in SDRG nor QMC calculations.     

\subsection{Spin correlations}
\label{subsec:C}

The ground state of the random Heisenberg chain with an even number $N$ of spins is a total-spin singlet and can be expressed in the VB basis. 
The approximate ground state resulting from the SDRG is described by a single set of bipartite valence bonds, $\ket{\psi_0}=\ket{v}$, with no bonds 
crossing each other, while the ground state projected out by the QMC method is a superposition of VB states,
\be
\ket{\psi_0}=\sum_v \alpha_v \ket{v},
\ee 
with non-negative coefficients $\alpha_v$ that are determined stochastically. In practice, the state by itself is not very useful
and one instead samples the contributions to the normalization $\langle \psi_0|\psi_0\rangle$ and accumulates the corresponding contributions
to expectation values of interest. 

%%%%%%%%%%%%%%%%%% FIG3 %%%%%%%%%%%%%%%%%%%%%%%%%%%%%%%%%%%%%%%%%%%%
\begin{figure}
\centerline{\includegraphics[width=5cm, clip]{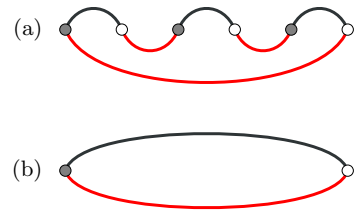}}
%\vskip-1mm
\caption{(Color online) Two different types of loop structures in the transition graph of the overlap between two VB states $\ket{v}$ 
(upper, black bonds) and $\ket{v'}$ (lower, red bonds). The gray and white sites belong to two different sublattices.
Type (a) indicates the case where $\ket{v}\neq \ket{v'}$ and type (b) is a single-bond structure with $\ket{v}=\ket{v'}$.
The correlation $C(\abs{i-j})$ for any pair of spins ($i$ and $j$) located in the same loop is a finite constant.}
\label{fig:loops_C}
%\vskip-3mm
\end{figure}
%%%%%%%%%%%%%%%%%%%%%%%%%%%%%%%%%%%

The matrix elements needed for computing the correlation function in the valence-bond basis are given by \cite{liang88,beach06}
\be
      \trip{v'}{{\bf S}_i\cdot{\bf S}_j}{v}
         =\begin{cases}
            \pm \frac{3}{4}\skp{v'}{v}, & (i,j),\\
                        0,   & (i)(j),  
          \end{cases} 
      \label{eq:matrix_SS}
\ee
where $(i,j)$ and $(i)(j)$ denote sites $i$ and $j$ belonging to the same loop and different loops, respectively. The sign in the one-loop case $(i,j)$ 
is positive for spins on the same sublattices and is negative otherwise. The overlap $\skp{v'}{v}$ between any two VB states is non-zero and
can be determined in terms of the total number of loops $N_\circ$
in the transposition graph,
\be
    \skp{v'}{v}=2^{N_\circ-N/2}
    \label{eq:overlap}
\ee
for $N$-spin VB states. In the SDRG case of a single bond configuration constituting the ground state
we have $\ket{v}=\ket{v'}$ and then the matrix for the two-spin operator in Eq.~(\ref{eq:matrix_SS}) is reduced to
\be
    \trip{v}{{\bf S}_i\cdot{\bf S}_j}{v} =
    \begin{cases}
        -\frac{3}{4},        & \text{if } i, j \text{ are connected by a bond,} \\
        0,                   & \text{if }  i, j \text{ are not connected.}
     \end{cases}
   \label{eq:matrix_SS_v}
\ee
In Fig.~\ref{fig:loops_C} we illustrate two different types of one-loop structures, corresponding to $\ket{v}\neq \ket{v'}$ and $\ket{v}=\ket{v'}$, 
in the transition graph of the overlap $\skp{v'}{v}$. A loop of type (b), which is the only kind of loop appearing in an overlap $\skp{v}{v}$ between
same states (as in the SDRG ground state), contains only two sites in different sublattices separated by an odd number of lattice spacings. A loop of 
type (a) can have an arbitrary even number of sites greater than two.
Thus the spin correlation function $C(r)$ for even $r$ is determined solely by loops of 
type (a). Since only loop-type (b) is present in the approximate SDRG ground state, in this case $C(r)=0$ for even $r$.

%%%%%%%%%%%%%%%%%%%%%%%%%%%%%%%%%%% FIG4 %%%%%%%%%%%%%%%%%%%%%%%
\begin{figure}
  \centerline{\includegraphics[width=7.5cm, clip]{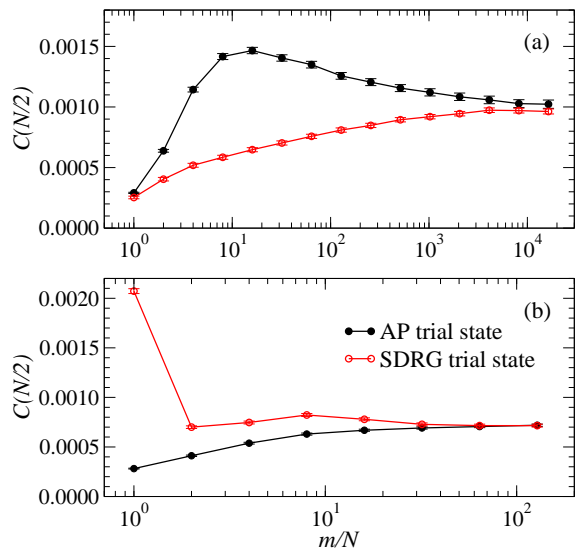}}
\caption{(Color online) Projection-power convergence of the disorder-averaged long-distance spin correlations in QMC calculations for
the (a) the random-$J$ system and (b) the random-$Q$ system, both with $d=1$ and $N=128$. Results are shown for two different trial states; 
the SDRG state obtained for each individual disorder realization and a translationally invariant amplitude-product state \cite{liang88,sandvik10} 
with bond-length ($l$) amplitude $h(l)=l^{-2}$.}
\label{fig:test}
\end{figure}
%%%%%%%%%%%%%%%%%%%%%%%%%%%%%%%%%%%%%%%%%%%%%%%%%%

Using the projector QMC method, for the random-$J$ model we have achieved
ground-state convergence for system sizes up to $N=144$ with $d=1$ in the
distribution (\ref{dist}) and up to $N=64$ for $d=2$, in each case using
between $10^4$ and $10^6$ disorder realizations to achieve sufficiently small
error bars on mean values. As an example of a convergence test, in
Fig.~\ref{fig:test}(a) we show results for the disorder-averaged spin
correlation function at the longest distance, $r=N/2$, of a random-$J$ system
with $d=1$. The two different trial states lead to values agreeing within error
bars, but with faster convergence observed with the SDRG states than a
translationally invariant amplitude-product state (where valence bonds are sampled 
according to probabilities given by products of bond amplitudes in the QMC procedure).\cite{liang88,sandvik10}  
For the random-$Q$ model the convergence is much faster
[Fig.~\ref{fig:test}(b)], and we have results for $N$ almost twice as large as
for the random-$J$ model (the results for the random-$Q$ model will be discussed in the next section).  
To speed up the equilibration of the QMC
calculations with high powers of $m$, an $m$-doubling procedure analogous to
the doubling procedure for the inverse temperature in Ref.~\onlinecite{sandvik02} was
used, where each simulation starts from $m=N$, after which $m$ is gradually
doubled by constructing an operator string of length $2m$ out of two consecutive
copies of the original string.

%%%%%%%%%%%%%%% FIG5 %%%%%%%%%%%%%%%%%%%%%%%%%
\begin{figure}
\centerline{\includegraphics[width=7.5cm, clip]{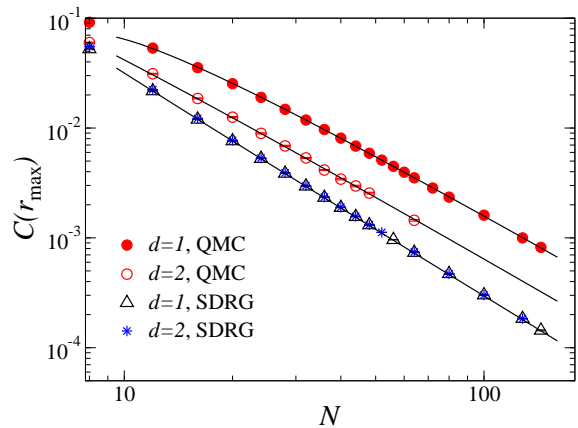}}
\vskip-1mm
\caption{(Color online) SDRG and QMC results for spin  correlations at the largest
distance of the random-$J$ model with disorder parameters $d=1,2$. The largest distance 
is $r_\text{max}=N/2$ for the QMC results, and $r_\text{max}=N/2-1$ for the SDRG data. 
The SDRG results have been fitted to the form $C(r)=\alpha r^{-2} + \beta r^{-4}$, with 
adjustable constants $\alpha$ and $\beta$. The form $C(r) \propto r^{-2}\ln^{1/2}(r/r_0)$ 
was used in fits to the QMC data.}
\label{figj}
\vskip-2mm
\end{figure}
%%%%%%%%%%%%%%%%%%%%%%%%%%%%%%%%%%%%%%%

%%%%%%%%%%%%%%%% FIG6 %%%%%%%%%%%%%%%%%%%%%%%%%%
\begin{figure}
\centerline{\includegraphics[width=7.5cm, clip]{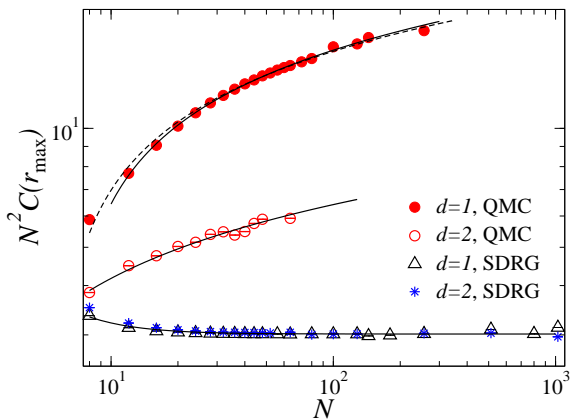}}
\vskip-1mm
\caption{(Color online) The correlation functions in Fig.~\ref{figj} multiplied by $N^2$ 
shows the presence of a multiplicative logarithmic correction to the $1/N^2$ scaling for the QMC results,
which increases with distance. The dashed line and the solid line for $d=1$ are both described by
the same correction form given in Eq.~(\ref{eq:log}), but with slightly different fitting parameters;
the dashed line for $d=1$ is the best fit for all data points, while the solid line (also shown in Fig.~\ref{figj}) 
is the best fit to the $N\ge 12$ data. The correction for the SDRG results shows a small enhancement for very small
$r$ and fast convergence to a constant, implying a small conventional subleading power-law correction.}
\label{fig:J_log}
\vskip-2mm
\end{figure}
%%%%%%%%%%%%%%%%%%%%%%%%%%%%%%%%%%%%%%%%%%%%%%%%%%%%%%

In Fig.~\ref{figj} results for the spin correlation function $C(r)$ at the largest distance $r=r_\text{max}$ is shown versus the chain length 
$N$ (even) and compared with SDRG results. The largest distance is $r_\text{max}=N/2$ for the QMC results, and $r_\text{max}=N/2-1$ for the SDRG 
data---there are no VBs of even length in the a 1D bipartite system, thus $C(N/2)=0$ in the SDRG case when $N$ is a multiple of $4$. 
The SDRG results follow the expected 
$r^{-2}$ decay for both distributions. We have here included a higher-power correction term to fit the data very closely also at short
distances, but the correction is very small and of no consequence for the longest distances shown. The QMC results clearly deviate from the
expected form and the deviations cannot be reasonably accounted for by any conventional correction. Previous works have not discussed how the 
asymptotic form is approached; it has merely been expected that, for long enough distances, results should approach the SDRG power law. At $r=N/2$ we 
observe clear deviations from $r^{-2}$ even for rather long distances. Remarkably, the data for both $d=1$ and $2$, and even for very small $r$, can 
be described by the form $C(r) \propto r^{-2}\ln^{\sigma_s}(r/r_0)$ with $0.3 \alt \sigma_s \alt 0.7$ and a scale parameter $r_0$ 
of order one.  To see the corrections more clearly, we multiply 
the results by $r^2$ in Fig.~\ref{fig:J_log} and show an excellent fit to the form with a multiplicative correction with $\sigma_s=0.5$ (in the
middle of the range of acceptable powers of the log factor),
\be
      r^2 C(r)=a\sqrt{\ln(r/r_0)},
      \label{eq:log}
\ee
here with $r=N/2$.
This multiplicative correction increases with $r$, and is clearly different from the additive correction term found in the numerical SDRG results
graphed in the same way.

%%%%%%%%%%%%%%%% FIG7 %%%%%%%%%%%%%%%%%%%%%%%%%%
\begin{figure}
\centerline{\includegraphics[width=7.5cm, clip]{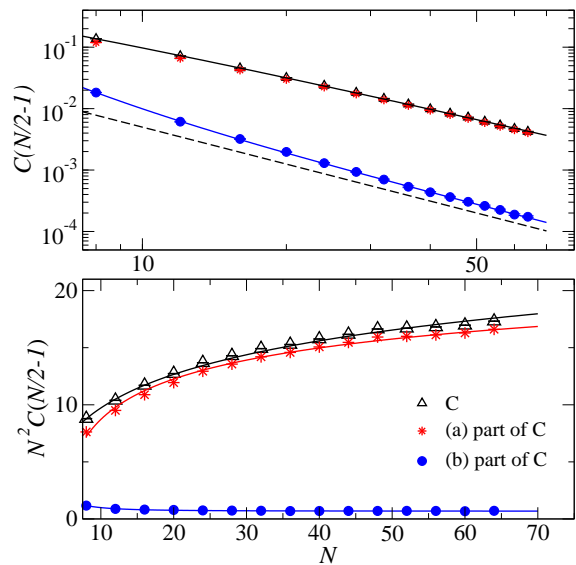}}
\vskip-1mm
\caption{(Color online) Upper panel: QMC results for the spin correlation function at the longest odd distance of
the random-$J$ chain with $d=1$ and its two components corresponding to the different types of loop structures in the overlap graph.
The dashed line indicates the inverse-square function $1/N^2$.
Lower panel: Corrections to the $1/N^2$ scaling made visible by dividing out the leading power law.
The logarithmic form in Eq.~(\ref{eq:log}) was used to fit the corrections
for the full correlation function and the (a)-part. The (b)-part of the correction has been fitted to a power-law form
$\alpha+\beta N^{-1}$.}
\label{fig:C_odd}
\vskip-2mm
\end{figure}
%%%%%%%%%%%%%%%%%%%%%%%%%%%%%%%%%%%%%%%%%%%%%%%%%%%%%%

As discussed above, the correlator $\e{{\bf S}_i\cdot{\bf S}_j}$ consists of two components corresponding to two different types of loop structures 
in the overlap graph. The QMC results shown in Fig.~\ref{figj} for the correlation at {\it even} $r$ are obtained entirely from loops of type (a), 
while the SDRG results are solely from the single-bond structure (b) and contain no even-$r$ correlations. This intriguing observation may explain 
the discrepancy between the SDRG and QMC results, and we explore this possibility next. 

To compare the QMC and SDRG results directly for the same distance, we have also calculated the spin correlation at the longest {\it odd} $r$ 
using QMC calculations at $d=1$ to examine the scaling of the two different $C(r)$-components originating from the loop structures in 
Fig.~\ref{fig:loops_C} (and we note here that these calculations were computed at a later stage and we did not go to the same large chain lengths as 
in the previous calculations focused only on $r=N/2$). As shown in Fig.~\ref{fig:C_odd}, the component (a) is the dominant part of the correlation 
and also exhibits multiplicative log correction to the inverse-square power-law scaling, consistent with the form in Eq.~(\ref{eq:log}). The 
component originating from loops of type (b) deviates from the $r^{-2}$ decay only at short distances and can be described by an additive subleading 
power-law, as in the SDRG case. All together, the correlation for odd $r$ is also well described by $C(r)\propto \sqrt{\ln(r/r_0)}/r^2$.
Since the loops of  type (a) are completely missing in the SDRG ground state, no multiplicative log correction of the origin identified here 
can be present within this approximation. This fundamental difference between the exact (QMC) and the single-VB SDRG ground states at least provides 
a technical explanation of why no log corrections appear within the SDRG, though we do not know the root cause for why the log is generated in 
the contribution of type (a) to the exact correlation function.

%%%%%%%%%%%%%%%%%%%%%%%%%%%%%%% FIG8  %%%%%%%%%%%%%%%%%%%%%%%%%%%%%%%%%

\begin{figure}[t]
\includegraphics[width=7.5cm, clip]{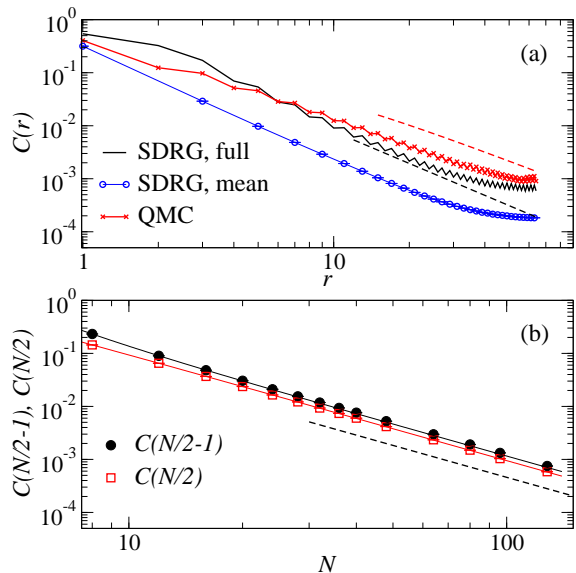}
\vskip-4mm
\caption{\label{fig:C_typ}
(Color online) (a) Full spin correlations for an $N=128$ random-$J$ chain with $d=1$, incorporating both the mean correlation and the typical 
correlation in the SDRG calculations. The results are compared with QMC results. The black and red dashed lines show, respectively, the pure
$1/N^2$ decay and that form modified by a multiplicative log of the same type used in the fit in Fig.~\ref{figj}. (b) The full SDRG spin correlation 
at the largest even and odd distances. The dashed line corresponds to the form $1/N^2$.}
\vskip-2mm
\end{figure}

%%%%%%%%%%%%%%%%%%%%%%%%%%%%%%%%%%%%%%%%%%%%%%%%%%%%%%%%%%%%%%%%%%%%%%

Our discussion about the SDRG treatment has so far focused on the mean spin
correlation, i.e., the dominant part of spin correlations originating from 
rare spin pairs that are strongly coupled by VBs. To incorporate the
correlations between the uncoupled singlets in the approximate SDRG ground state,
we need to keep track of weak effective couplings in the RG procedure that
induce correlations between typical pairs of spins.\cite{fisher94,fisher98}
For example, at some RG step of energy scale $\Omega$, a pair of spins ${\bf
S}_j,\,{\bf S}_k$ with the strongest coupling of strength $\Omega$ is
decimated. The spin ${\bf S}_j$ will become strongly correlated with ${\bf
S}_k$ and form a singlet pair, but is only weakly correlated to its other
neighbor, say ${\bf S}_i$, which is a spin that survives at this decimation 
step.  The correlation between this just-decimated spin ${\bf S}_j$ and its 
weak-side neighbor ${\bf S}_i$ can be obtained by first-order perturbation
theory,\cite{fisher94,fisher98} yielding
\be
     \e{{\bf S}_i\cdot{\bf S}_j} \approx \frac{\tilde{J}_i}{\Omega} \e{{\bf S}_j\cdot{\bf S}_k},
     \label{eq:C_typ_eq}
\ee
where $\tilde{J}_i$ is the (effective) coupling between $i$ and $j$ at energy scale $\Omega$, 
and $\abs{\e{{\bf S}_j\cdot{\bf S}_k}}\approx3/4$ for the strongly correlated spin pair.
As pointed out in Ref.~\onlinecite{fisher94}, the distribution of the
logarithmic couplings $\zeta_i\equiv \ln(\Omega/\tilde{J}_i)$ becomes broader and broader
under renormalization, and the weak correlations generated by Eq.~(\ref{eq:C_typ_eq}), which
constitute the typical correlations, then decay exponentially with the distance as in Eq.~(\ref{eq:C_typ}). 

In Fig.~\ref{fig:C_typ} we incorporate both the mean and the typical correlation contributions in the perturbative 
SDRG calculations. We here graph the results versus the distance $r$ in a chain of length $L=128$, instead of investigating
the scaling at the largest distance versus $N$. The two ways of analyzing correlation functions should give the same
functional form, but with different prefactors because of the elevated amplitude of the correlations close to $r_{\rm max}$.
While the inclusion of the typical correlations brings the result significantly closer to the non-perturbative (numerically exact) 
QMC result (including even the expected even-odd oscillations \cite{hoyos07}), the asymptotic decay is still, as expected, governed 
by the mean spin correlation, thus following the inverse-square law. The QMC data deviate from this form but can be well decribed by 
including the multiplicative log (though, as expected, this is not as clear as in the previous analysis of the system-size 
dependence). We do not see any apparent way to modify the SDRG method to generate the multiplicative log correction seen in the 
QMC calculation; likely it originates from a mechanism which is beyond the capability of a renormalization scheme such as the SDRG.

As mentioned above, a multiplicative log correction is present for the clean system (with $\sigma_s=1/2$, which we also 
use in the fits shown although other exponents close to this value also work well), but the marginal operator responsible 
for it is not expected to play any role in a strongly disordered system. Logs produced by perturbative disorder have been
demonstrated in certain systems without marginal operators in the clean limit,\cite{ristivojevic12} and it has also been argued 
that the correlations in the strongly-disordered XX chain are affected by a log correction,\cite{igloi00} and this would again
not be related to any marginal operator in the clean limit. We are not aware of any previous suggestions of log corrections 
in the random exchange Heisenberg chain, but we regard the numerical evidence presented above as very strong.

\subsection{Dimer correlations}
\label{sec:D}

%%%%%%%%%%%%%%%%%%%%%%%%%%%%%%% FIG9  %%%%%%%%%%%%%%%%%%%%%%%%%%%%%%%%%
\begin{figure}[ht]
\includegraphics[width=5.5cm, clip]{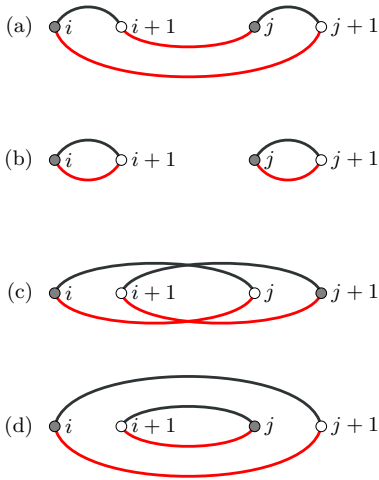}
\vskip-4mm
\caption{\label{fig:D_loops}
(Color online) Four types of loop structures in the transition graph of the overlap $\skp{v'}{v}$ that contribute to
dimer correlations. The black and red bonds correspond to $\ket{v}$ and $\bra{v'}$, respectively. The gray and white
sites belong to two different sublattices. 
Types (a) and (c) are absent in the SDRG ground state since $\ket{v}=\ket{v'}$ and only non-crossing bonds are
present, while these cases do contribute in QMC simulations where
type (c) with two loops $(i,j)(j+1,i+1)$ can be realized using more than 4 spins connected with
non-crossing bipartite bonds and $\ket{v}\neq\ket{v'}$.}
%\vskip-2mm
\end{figure}
%%%%%%%%%%%%%%%%%%%%%%%%%%%%%%%%%%%%%%%%%%%%%%%%%%%%%%%%%%%%%%%%%%%%%%

Now we turn to four-spin correlations defined in Eq.~(\ref{eq:D_def}). 
For a general case, the matrix elements $\trip{v'}{B_i B_j}{v}$ have finite values
for four different situations depending on the loop-structure in the transition graph
of the overlap $\skp{v'}{v}$.\cite{beach06,tang11b} Using a notation where sites enclosed by
$()$ belong to the same loop, the four types of site-loop-structures are: (a) $(i, i+1, j, j+1)$;
(b) $(i, i+1)(j,j+1)$; (c) $(i, j)(j+1, i+1)$; (d) $(i+1, j)(i, j+1)$. We illustrate these cases 
for four spins in Fig.~\ref{fig:D_loops}. A complete formula for evaluating $\trip{v'}{B_i B_j}{v}$ 
can be found, e.g., in Ref.~\onlinecite{tang11b}.  

\begin{figure}
\centerline{\includegraphics[width=7.5cm, clip]{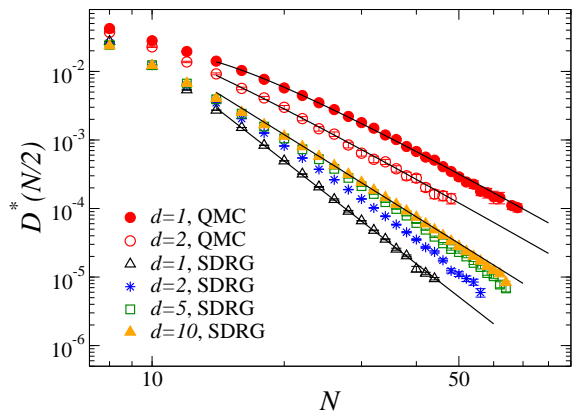}}
%\vskip-1mm
\caption{(Color online) SDRG and QMC results for dimer correlations of the random-$J$ model with different 
disorder parameters $d$. The fitting forms are $D^*(r) = \alpha r^{-4} + \beta r^{-5}$ (SDRG) and $D^*(r) \propto r^{-4}\ln(r/r_0)$.}
\label{fig:D_J}
\end{figure}

We define the staggered dimer correlation function using the definition of $D(r)$ in
Eq.~(\ref{eq:D_def}) as 
\begin{equation}
D^*(r)=[D(r)-\hbox{$\frac{1}{2}$}D(r-1)-\hbox{$\frac{1}{2}$}D(r+1)](-1)^r,
\label{eq:D_star}
\end{equation}
and shows results for $r=N/2$ versus $N$ in Fig.~\ref{fig:D_J}. The SDRG results can
be fitted to the form $D^*(r) = \alpha r^{-4} +\beta r^{-5}$, with constants $\alpha$ and $\beta$ depending 
on $d$. The slower decay of the QMC data can again not be reasonably explained by conventional corrections but are very well accounted
for by a multiplicative log, $D^*(r) \propto r^{-4}\ln^{\sigma_d}(r/r_0)$, with $\sigma_d \approx 1$ (good fits require $0.5 \alt \sigma_d \alt 1.5$) 
and only the scale parameter $r_0$ depending on $d$. 

\begin{figure}
\includegraphics[width=8cm, clip]{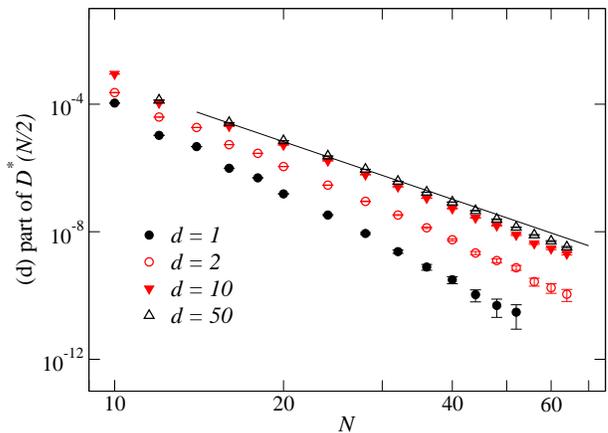}
\vskip-2mm
\caption{\label{fig:D_loop_d}
(Color online) Contribution to the dimer correlation from the loop case labeled (d) in Fig.~\ref{fig:D_loops}
at the largest distance for the random $J$ chain with different disorder strength $d$.
Each data point was obtained by averaging over more than $10^{4}$ disorder realizations. The solid line indicates a power-law
decay as $N^{-6}$, showing consistency with an asymptotic distance dependence $\propto r^{-6}$ for large $d$.}
\vskip-2mm
\end{figure}

It is tempting to interpret $D^*(r)$ as the square of $C(r)$, but there is nothing obvious in the definition of the dimer correlation 
function or its valence-bond estimator to suggest such a relationship. 
In the case of the single non-crossing VB state resulting from the SDRG procedure, there are two contributions to $D(r)$, 
from cases (b) and (d) in Fig.~\ref{fig:D_loops}, out of four in a completely general state.\cite{beach06} We find that the contributions from 
two nearest-neighbor bonds completely dominate $D(r)$ and $D^*(r)$ in the SDRG ground state, with contributions from longer bonds decaying 
with a higher power as shown in Fig.~\ref{fig:D_loop_d}. In the QMC calculations the full loop representation of the correlations come into 
play \cite{beach06} and the interpretation of the different contributions to $D(r)$ is less clear-cut.

%%%%%%%%%%%%%%% FIG10 %%%%%%%%%%%%%%%%%%%%%%%%%
%%%%%%%%%%%%%%%%%%%%%%%%%%%%%%%%%%%%%%%
\null\vskip5mm

%%%%%%%%%%%%%%%%%%%%%%%%%%%%%%% FIG11  %%%%%%%%%%%%%%%%%%%%%%%%%%%%%%%%%

%%%%%%%%%%%%%%%%%%%%%%%%%%%%%%%%%%%%%%%%%%%%%%%%%%%%%%%%%%%%%%%%%%%%%%

\section{Amorphous valence-bond solid}
\label{sec:avbs}

The clean $Q$ model with six-spin interactions is VBS ordered and when combined with the Heisenberg exchange $J$ it undergoes a transition 
to the standard critical antiferromagnet at $J/Q\approx 6$.\cite{tang11a,banerjee10} The dimerization transition is in the same universality class as that
in the well-studied $J_1$-$J_2$ Heisenberg chain.\cite{affleck85,nomura92,eggert96} An important question is how disorder affects such a transition 
and the VBS state. In the latter, one can expect an AVBS with alternating domains of the two different dimerization patterns (which differ by a translation 
of one lattice unit), and a simple valence-bond picture suggests that the domain walls between these domains should contain $S=1/2$ spin degrees of 
freedom---localized spinons---corresponding to long valence bonds between different domain walls as illustrated in Fig.~\ref{fig1}. 

Localized spinons were recently observed in a study combining SDRG,
variational, and DMRG calculations for the $J_1$-$J_2$ chain with disorder
added at the special Majumdar-Ghosh (MG) point $J_2=J_1/2$, where the exact
ground state is a doubly-degenerate short-bond VBS.\cite{lavarelo13}  For a
certain type of correlated disorder satisfying the conditions underlying the MG
exact ground state, an Anderson-type spinon localization mechanism was
identified at a critical disorder strength. If the MG condition is violated,
the SDRG procedure can generate mixed ferromagnetic and antiferromagnetic
couplings, leading to a partially polarized ferromagnet, as in the ``large
spin'' phase first identified in Ref.~\onlinecite{westerberg97}. 

Unlike the Anderson-localization transition found in Ref.~\onlinecite{lavarelo13}, in our model there is no special condition precluding 
a transition between the VBS and an AVBS with spinons at infinitesimal disorder strength, following the Imry-Ma arguments \cite{imry75} 
as applied to gapped Mott insulators turning into gapless Anderson insulators.\cite{shankar90,pang93} Thus, for arbitrarily weak disorder, in an infinite 
chain there will be some regions favoring one ordering pattern (singlets of even or odd bonds) and some other regions favoring the other pattern. The
typical size of the domains diverges as the disorder strength vanishes. The resulting state at finite disorder should be similar to the one 
with domain-wall spinons found in Ref.~\onlinecite{lavarelo13}. However, in that case it was argued that strong disorder (small VBS domains) will  
lead to some effective ferromagnetic spinon-spinon couplings and a partially polarized state. As no ferromagnetic couplings are generated in the random-$Q$ model, 
this system offers opportunities to study a generic singlet AVBS where the size of the VBS domains can be tuned from infinity in the clean system down to
small lengths where the picture of VBS domains and domain walls breaks down. In a $J$-$Q$ model, the ratio $J/Q$ further offers the possibility to also tune 
the strength of the dimer order in the domains and the (related) spinon localization length. Here we focus on the $Q$ model, which has strong VBS order in 
the clean limit, adding disorder according to the distribution (\ref{dist}).  

%%%%%%%%%%%%%%%%%% FIG12 %%%%%%%%%%%%%%%%%%%%%%%%%%%%%%%%%
\begin{figure}
\centerline{\includegraphics[width=7.5cm, clip]{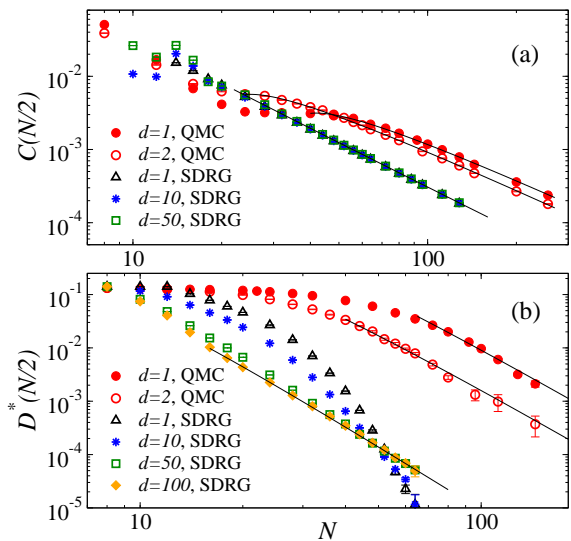}}
\vskip-2mm
\caption{(Color online) SDRG and QMC results for the spin (a) and dimer (b) correlations in the random-$Q$ model. The functional
forms fitted to the data (curves shown) are the same as in the corresponding cases in Fig.~\ref{figj}.}
\label{figq}
\end{figure}
%%%%%%%%%%%%%%%%%%%%%%%%
\null\vskip2mm

%%%%%%%%%%%%%%%% FIG13 %%%%%%%%%%%%%%%%%%%%%%%%%%
\begin{figure}
\centerline{\includegraphics[width=7.5cm, clip]{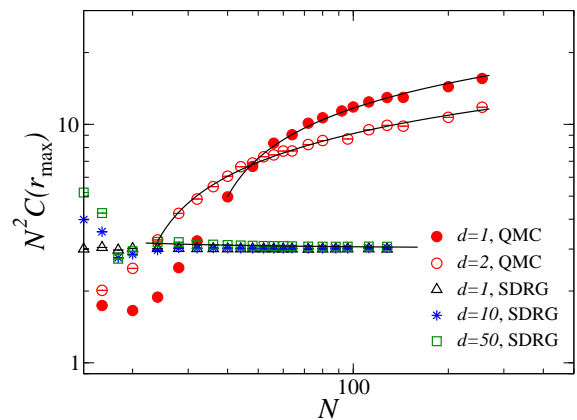}}
\vskip-1mm
\caption{(Color online) Corrections to the $1/r^2$ scaling of the spin correlations of the random-$Q$ model, made
visible by dividing the data in Fig.~\ref{figq}(a) at $r=N/2$ by the leading power-law.}
\label{fig:Q_log}
\vskip-2mm
\end{figure}
%%%%%%%%%%%%%%%%%%%%%%%%%%%%%%%%%%%%%%%%%%%%%%%%%%%%%%

Looking at the QMC spin correlations of the $Q$-model graphed in
Fig.~\ref{figq}(a), there is first a rapid decay, followed by a plateau, after
which the asymptotic decay is consistent with the same $r^{-2}$ form with
multiplicative log correction found in the random-$J$ model. The deviations
from the leading power-law behavior is again made visible by multiplying by $N^2$
in Fig.~\ref{fig:Q_log}.

The SDRG results show a different short-distance behavior, but again the
asymptotic form is $r^{-2}$. It is not surprising that the SDRG method cannot
fully capture the correlations at short distance, since it is expected to
become accurate (in an RS state) only gradually as the process flows toward the
RS fixed point. It is nevertheless interesting to see that the behavior is
quite different from that in the random-$J$ model, where a behavior very close
to $r^{-2}$ sets in already at the shortest distances. Both the SDRG and QMC
calculations support the notion that there are localized spinons in the AVBS,
which form a gapless random spin sub-system governed by the RS fixed point.

Next, we analyze the dimer correlations in Fig.~\ref{figq}(b). In the QMC
results the almost flat plateaus at short distances reflect the presence of
ordered VBS segments, with a typical length (size of the plateau) which depends
on the disorder distribution. Beyond the plateau, the behavior is consistent
with $r^{-4}$ decay with a multiplicative log correction, again fully
consistent with the behavior of the random-$J$ model.  The SDRG results at very
large $d$ show $r^{-4}$ decays with a correction in the form of an additive
higher power. For smaller $d$ the decay appears faster, but the behaviors for
different values of $d$ indicate that this is only a cross-over to an $r^{-4}$
decay with a very small amplitude.
The VBS domains of the AVBS for $d=1,2$ are partially captured by the SDRG, though
there is no flat plateau, merely a slower initial decay.

The scale parameter $r_0$ in the log factor $\ln^\sigma(r/r_0)$ describing the QMC correlation
functions is of the order $10$ in the random-$Q$ model for all the cases (spin and dimer correlations), i.e., much 
larger than the values of order $1$ in the random-$J$ model. This naturally reflects an effective renormalized 
lattice spacing in the sub-system of localized spinons, which forms the RS state.

We comment on the difficulties in observing the expected asymptotic $r^{-4}$ decay in the SDRG calculations for
random-$Q$ model with small $d$ in Fig.~\ref{figq}(b). Considering the 3-stage evolution of the energy scale in the RG process
demonstrated by Fig.~\ref{fig:SDRG_gaps} for $d=1$, the dimer correlations should also be sensitive to these
three SDRG stages. Unfortunately, due to the very small values of the correlation functions and associated
large relative statistical fluctuations [$D^*$ defined in Eq.~(\ref{eq:D_star}) contains positive and negative contributions
which almost cancel each other], we are only able to compute the dimer correlation to high precision in short
chains, typically using at least $10^{10}$ random coupling samples. Therefore, we only reach the early RG stage,
which shows a fast decay for small $d$. With larger $d$, the final stage truly reflecting the RS ground state
can be reached. It can be noted here again that the spin correlations are only sensitive to the bond-length
distribution, which converges relatively fast, while the dimer correlations depend on long-distance bond-bond
correlations. 

\section{Discussion}
\label{sec:discussion}

Our SDRG and QMC results show consistently that both the random-$J$ and the random-$Q$ models are asymptotically governed
by the RS fixed point. Thus, in a $J$-$Q$ model, we do not expect any phase transition as a function of the ratio $J/Q$, unlike the clean 
system where there is a dimerization transition of the same universality class as in the $J_1$-$J_2$ Heisenberg chain.\cite{tang11a,banerjee10} 
For weak disorder and $J/Q$ in the neighborhood of its critical value in the clean system, there should be interesting combined effects of 
the critical fluctuations and RS physics. Although there is no phase transition in the sense of asymptotic, the AVBS can still be considered as 
a state of matter different from the Heisenberg-RS, because it possesses a length-scale---that of VBS domains---which is not present (or, more
precisely, it is of order the lattice spacing) at the RS fixed-point alone, but which can be made arbitrarily large by tuning
interactions in the AVBS state.

The RS fixed point is exact for the SDRG scheme, but our findings of log corrections suggest that systems treated without approximations flow 
to this point (under, e.g., increase of the system size or lowering of the energy scale in an infinite system) slower than expected. The same 
leading power laws and log corrections consistently describe the correlations in the random-$J$ and random-$Q$ models with different disorder 
distributions, demonstrating a robust universality of the log exponents characterizing the RS phase. We have shown explicitly that the SDRG
method is fundamentally incapable of producing the log correction to the mean correlation function, because in the unbiased QMC treatment
it originates in the VB basis from a loop structure that is never generated within the SDRG.

The physics of the VBS and AVBS also applies to spin chains coupled to phonons. In the classical limit, any spin-phonon coupling leads to dimerization 
(the spin-Peierls distortion), while at finite phonon frequency a critical coupling is required.\cite{sandvik95,uhrig96,suwa15} The 
relationship between this transition and that in the $J_1$-$J_2$ chain is well established \cite{uhrig98,weisse99} and the $J$-$Q$ model provides 
an alternative to access the same physics.\cite{tang11a,banerjee10} The AVBS state we have identified and characterized here should be relevant 
to quasi-one-dimensional spin-phonon materials, e.g., CuGeO$_3$ \cite{hase93} and TiOCl.\cite{zhang14} RS scaling due to localized 
spinons should be detectable using NMR, and it would then be desirable to also calculate temperature dependent magnetic properties. It may also be
possible to study AVBS-related disorder effects in dimerized phases of a trapped-ion system.\cite{bermudez15}

\begin{acknowledgments}

We would like to thank Leon Balents, Ferenc Igl\'oi, and Nicolas Laflorencie for useful discussions.
The work of  CWK and YCL was supported by Ministry of Science and Technology, Taiwan, under Grants No.~105-2112-M-004-002, 
104-2112-M-004-002, 101-2112-M-004-005-MY3, and by NCTS (Taiwan). YRS and DXY acknowledge support from Grants
NBRPC-2012CB821400, NSFC-11275279, NSFC-11574404, NSFG-2015A030313176, and Special Program for Applied Research on Super 
Computation of the NSFC-Guangdong Joint Fund (the second phase). AWS was supported by the NSF under Grant No.~DMR-1410126 and 
by the Simons Foundation.

\end{acknowledgments}

\appendix

\section{SDRG for the random-$Q$ interaction}
\label{app:sdrg}

\begin{figure*}[t]
\includegraphics[width=14cm, clip]{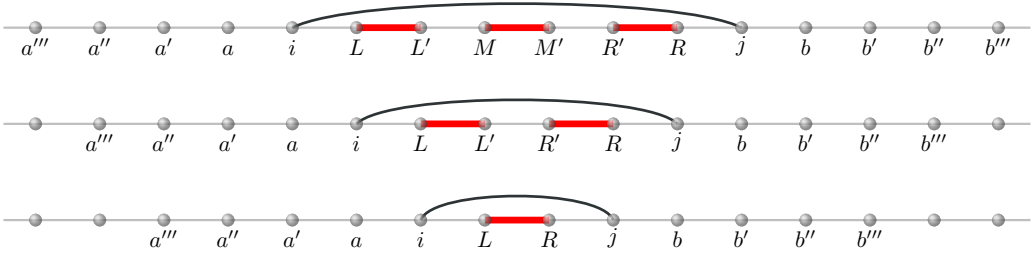}
\vskip-3mm
\caption{\label{fig:SDRG_Q3}
(Color online) SDRG rules for the $Q_3$ chain. The thick bonds indicate the strongest coupling (a $Q_3$-coupling [top], a $Q_2$-coupling [middle] or a 
$J$-coupling [bottom]) to be decimated; each pair of spins connected to the thick bonds forms singlet. The decimation procedure leads to an effective coupling 
between sites $i$ and $j$.}
\end{figure*}

Here we describe the SDRG procedure for the random $Q$-chain with six-spin interactions,
\be
     H_{Q_3}=-\sum_i Q_i P_{i,i+1} P_{i+2,i+3} P_{i+4,i+5},
\ee
where $Q_i>0\;\forall i$ and $P_{ij}=1/4-{\bf S}_i\cdot{\bf S}_j$ is a singlet projector acting on spins ${\bf S}_{i}$ and 
${\bf S}_{j}$. During the decimation process, effective two-site interactions ($J$-terms) $-JP_{i,j}$ and four-site interactions 
($Q_2$-terms) $-QP_{i,i+1}P_{i+2,i+3}$ will be generated; therefore, the RG procedure described below is valid for a general 
random $J$-$Q_3$-chain obtained by combining $H_{Q_3}$ and the Heisenberg model,
\begin{eqnarray}
   H_{J\hbox{-}Q_3} & = & -\sum_i J_i P_{i,i+1} \\
               && -\sum_i Q_i P_{i,i+1} P_{i+2,i+3} P_{i+4,i+5},\nonumber
\end{eqnarray}
and also for the $J$-$Q_2$ chain where the $Q_3$ terms are replaced by $Q_2$ terms. In our discussion below, we use the notation 
introduced in Fig.~\ref{fig:SDRG_Q3} to indicate the spins involved in an RG decimation.

%%%%%%%%%%%%%%%%%%%%%%%%%%%%%%% FIG14  %%%%%%%%%%%%%%%%%%%%%%%%%%%%%%%%%
%%%%%%%%%%%%%%%%%%%%%%%%%%%%%%%%%%%%%%%%%%%%%%%%%%%%%%%%%%%%%%%%%%%%%%

\subsubsection{$Q_3$-decimation}
Consider a 6-spin coupling ($Q_3$-term) with coupling strength $Q_0$ which is the dominant term at some stage of the RG, i.e.
$\Omega=Q_0$ sets the RG energy scale. The ground state of the associated part of the Hamiltonian, 
\be
   H_0 = -Q_0 P_{L,L'} P_{M, M'} P_{R', R},
    \label{eq:Q3_H0}
\ee
is a 6-spin singlet and the excited states are 63-fold degenerate (which in the basis of bond singlets and triplets simply follows from 
the fact that at least one of the singlet projectors give $0$ unless they all act on singlets); the energy gap between the ground state 
and the excited multiplet is $\Omega=Q_0$. Different from the SDRG process on the standard Heisenberg chain, under the action of the RG 
there are here more than two possible neighboring terms of $H_0$ in the random $Q$-chain. Below we consider the effect of all of 
these cases of interaction terms perturbatively. Typically there are several options for which terms to select for the
perturbative treatment to generate the new couplings, and one can sum the contributions from all of them or select just the
one generating the largest contribution. We will discuss this aspect of our practical implementation further below in 
Sec.~\ref{sec:imp}, after first discussing how to generate the perturbative coupling for all possible cases.

~\\
\noindent
{\bf Case (i):} 
The neighboring terms form a pair of $Q_3$-couplings with outermost bonds at $[i,L]$ and $[R,j]$ where $i$ and $j$ are nearest neighboring 
sites to the left and right spins, ${\bf S}_L$ and ${\bf S}_R$, respectively, of the 6-spin segment to be decimated:
\be
    H_1=-Q P_{i,L}P_{L',M}P_{M',R'}-Q' P_{L',M}P_{M',R'} P_{R,j}.
\ee
To obtain a nonzero coupling joining together the spins on both sides of $[L,R]$ including
${\bf S}_i$ and ${\bf S}_j$, we need to use second-order perturbation theory
since the first-order contribution vanishes.
This leads to an effective coupling: 
\be
    \tilde{J}_{ij}=\frac{Q Q'}{32 \Omega}.
\ee

~\\
\noindent
{\bf Case (ii):}
The neighboring $Q_3$-terms include outermost bonds at $[a',a]$ and/or $[b, b']$ (see Fig.~\ref{fig:SDRG_Q3}), with
the possible effective Hamiltonians being
\be
    H_1=-Q P_{a',a} P_{i,L} P_{L',M}-Q' P_{M',R'} P_{R,j}P_{b,b'},
\ee
or
\be
    H_1=-Q P_{a',a} P_{i,L} P_{L',M}-Q' P_{L',M} P_{M',R'} P_{R,j},
\ee
or
\be
    H_1=-Q P_{i,L} P_{L',M} P_{M',R'}-Q' P_{M',R'} P_{R,j} P_{b,b'}.
\ee
For all these cases we obtain an effective coupling between the spins ${\bf S}_i$ and ${\bf S}_j$:
\be
   \tilde{J}_{ij}=\frac{Q Q'}{32 \Omega},
\ee
and, in addition, the operator $-P_{a',a}$ or $-P_{b,b'}$ in $H_1$, which is outside the decimated region $[i,\,j]$, is converted 
to a $J$-bond of strength $Q/16$ or $Q'/16$ to first order in perturbation theory. 

~\\
\noindent
{\bf Case (iii):}
The pair of neighboring $Q_3$-terms includes one $Q_3$-coupling with the outermost bond $[i,L]$ (or $[R,j]$) and one $Q_3$-coupling with the innermost 
bond $[R,j]$ (or $[i,L]$), i.e.,
\be
   H_1=-Q P_{i,L} P_{L',M} P_{M',R'}-Q' P_{R,j} P_{b,b'} P_{b'',b'''}, 
\ee 
or 
\be
   H_1=-Q P_{a''',a''} P_{a',a} P_{i,L} -Q' P_{L',M} P_{M',R'} P_{R,j}.
\ee
To second order we obtain
\be
   \tilde{J}_{ij}=\frac{Q Q'}{32 \Omega}\,,
\ee
for an effective coupling between sites $i$ and $j$, and in the process to first order the operator $-P_{a''',a''} P_{a',a}$ 
or $-P_{b,b'} P_{b'',b'''}$ is converted to a 4-spin $Q_2$-term of strength $Q/4$ or $Q'/4$.

\noindent
~\\
{\bf Case (iv):}
There are certain pairs of neighboring $Q_3$-terms that do not contribute to effective couplings between sites $i$ and $j$, 
for example, when one of the $Q_3$-term contains no bond at $[i,L]$ and $[R,j]$, such as \\
~~$-QP_{a,i}P_{L,L'}P_{M,M'}$,\\
~~$-QP_{a'',a'}P_{a,i}P_{L,L'}$,\\ 
~~$-QP_{M,M'}P_{R',R}P_{j,b}$,\\
~~$-QP_{R',R}P_{j,b}P_{b',b''}$.\\ 
There are also cases of operators containing bonds $[i,L]$ or $[R,j]$ but still give zero contribution 
to $\tilde{J}_{ij}$, such as:
\be
   H_1=-Q P_{a''',a''} P_{a',a} P_{i,L} - Q' P_{R, j} P_{b,b'} P_{b'',b'''}.
\ee

When part of a neighboring $Q$-term is decimated (i.e., the bonds in the region $[i,\,j]$ are removed) as explained
above, the surviving part of the $Q_3$ term outside the decimation region will be converted to a 4-spin $Q_2$-coupling 
or a two-spin $J$-coupling. The strength of the surviving part, obtained in this case via first-order perturbation 
theory, depends on the location of the decimated part in the region $[i,\,j]$, where three two-spin singlets at $[L, L'],\,[M, M']$ 
and $[R', R]$ are formed. The general rule is as follows: in the decimated region a bond-operator located between two sites that do 
not form a singlet (sites on which no operator in the dominant $H_0$-term acts), e.g., between sites $L'$ and $M$, will reduce 
the strength of the surviving part by a factor $1/4$ (and, accordingly, the strength of the surviving part of an operator with two 
such bonds will include a factor $1/16$), while a decimated bond-operator on a singlet will not modify the strength of surviving 
part. For example, a neighboring $Q_3$-term such as $-Q P_{a''',a''} P_{a',a} P_{i,L}$ will be converted to a $Q_2$-term 
$-(Q/4)P_{a''',a''} P_{a',a}$, while a $Q_3$-term such as $-QP_{a,i}P_{L,L'}P_{M,M'}$ will be converted to $-QP_{a,i}$. 
This truncation rule is applied in cases (ii), (iii), and (iv). 

~\\
As is apparent from the above discussion, during the RG procedure effective $J$-terms and $Q_2$-terms will be generated in the
system; they are either bonds truncated from perturbative $Q_3$-terms, or those effective $J$-couplings generated 
between the neighboring spins of a decimated $Q_3$-term. 
These $J$- or $Q_2$-couplings will also generate effective $\tilde{J}_{ij}$ when they become 
perturbative terms to a dominant $Q_3$-term [cf. cases (i), (ii) and (iii) above].
We note the following cases:

~\\
\noindent
{\bf Case (v):}
One $Q_3$-coupling and one $J$-coupling as the perturbative terms, e.g.,
\be
   H_1=-Q P_{i,L} P_{L', M} P_{M', R'}-J P_{R,j} ,
\ee
where $R$ and $j$ may be arbitrarily distant sites. Up to second order, we obtain an effective coupling between 
sites $i$ and $j$:
\be
      \tilde{J}_{ij}=\frac{Q J}{32 \Omega}.
\ee
Similarly, for the case
\be
   H_1=-Q P_{i,L} P_{L', M} P_{M', R'}-Q' P_{R,j}P_{a,a''} ,
\ee
we obtain
\be
      \tilde{J}_{ij}=\frac{Q Q'}{32 \Omega}.
\ee

~\\
\noindent
{\bf Case (vi):}
One $Q_2$-coupling and one $J$-coupling as the perturbative terms, e.g.,
\be
   H_1=-Q P_{i,L} P_{L', M}-J P_{R,j}.
\ee
Exact diagonalization of the block with $H_0$ and $H_1$ shows the ground state of the block
is four-fold degenerate, indicating zero couplings between $i$ and $j$; 
Also with a perturbation such as
\be
   H_1=-J P_{i,L}-J' P_{R,j},
\ee  
we obtain no coupling between $i$ and $j$. These perturbative terms are simply removed 
(see Sec.~\ref{sec:unpaired} for further discussion of rare special cases where all effective
couplings vanish).

\subsubsection{$Q_2$-decimation}

Consider a $Q_2$-term which is the dominant term at some stage of the RG.
The ground state of the associated part of the Hamiltonian,
\be
   H_0 = -Q_0 P_{L,L'} P_{R', R},
   \label{eq:Q2_H0} 
\ee
is a 4-spin singlet and the excited states are 15-fold degenerate; the energy
gap between the ground state and excited multiplets is $\Omega=Q_0$.
Below we list the perturbative terms which generate effective couplings between
the neighboring spins ${\bf S}_i$ and ${\bf S}_j$:

~\\
\noindent
{\bf Case (i):}
The neighboring perturbative terms constitute a pair of $Q$-couplings ($Q_3$ or
$Q_2$) with strength $Q$ and $Q'$, and at least one of the coupling does not contain 
the innermost bond at $[i,L]$ or $[R,j]$ (which is equivalent to a $J$-coupling at $[i,L]$ or $[R,j]$). 
For this case, to second order we obtain an effective coupling
\be
     \tilde{J}_{ij}=\frac{Q Q'}{8\Omega}.
\ee 
between sites $i$ and $j$. The bonds outside the region $[i,j]$ will be converted to $Q_2$ or $J$-couplings, 
following the truncation rules discussed after Case (iv) of the $Q_3$-decimation.  The same result holds when one coupling of
the perturbative terms is a $J$-coupling and one coupling is a $Q_3$ or $Q_2$-coupling that is not equivalent 
to a $J$-term.

~\\
\noindent
{\bf Case (ii):}
The dominant $Q_2$-term $H_0=-QP_{L,L'}P_{R',R}$ is embedded in
a $Q_3$-coupling $-QP_{i,L}P_{L',R'}P_{R',j}$. To first order, the effective
coupling between $i$ and $j$ is 
\be
    \tilde{J}_{ij}=\frac{Q}{16}. 
\ee

\subsubsection{J-decimation}

Here we consider the two cases when a $J$-term $H_0=-\Omega P_{L,R}$ is the strongest coupling at some step of RG. 

~\\
\noindent
{\bf Case (i):}
One $Q$-term such as 
\be
   H_1=-Q\,P_{i, L}\,P_{R, j},
\ee
or
\be  
  H_1=-Q\,P_{i, L}\,P_{R, j}\,P_{b, b'},
\ee
or
\be
  H_1=-Q\,P_{a',a}\,P_{i, L}\,P_{R, j},
\ee  
as a perturbation.
To first order, we obtain
\be
  \tilde{J}_{ij}=\frac{Q}{4}.
\ee

~\\
\noindent
{\bf Case (ii):}
The perturbative terms constitute a pair of $J$-couplings:
\be
   {H}_1= -J P_{i,L} -J' P_{R,j}. 
\ee 
This is an RG decimation for the standard Heisenberg chain, in which
an effective coupling
\be
     \tilde{J}_{ij}=\frac{J J'}{2\Omega}.
\ee
is generated. Similarly, 
if the perturbative terms constitute a pair of $Q$-couplings such as
\be
  {H}_1= -Q P_{a''',a''}P_{a',a} P_{i,L} -Q' P_{R,j} P_{b,b'} P_{b'',b'''},
\ee
the effective coupling is
\be
   \tilde{J}_{ij}=\frac{Q Q'}{2\Omega}.
\ee
Also, with one $Q$-coupling and one $J$-coupling as a perturbation, we obtain
\be
  \tilde{J}_{ij}=\frac{Q J}{2\Omega}. 
\ee

\subsection{Implementation}
\label{sec:imp}

In our numerics, we use the {\it maximum rule} in the recursion relations
for generating the effective couplings ($J$ or $Q$ couplings):
\be
   \ln(\lambda)=\ln(\lambda_1+\lambda_2) \approx \max \bigl[\ln (\lambda_1),\,\ln (\lambda_2)  \bigr],
   \label{eq:maximum_rule}
\ee 
where $\lambda_1$ and $\lambda_2$ are bonds connecting the same group of spins.
We have compared the results with those obtained by using the {\it sum rule},
in which we sum the newly generated coupling and the pre-existing coupling to
obtain the effective coupling. We have found no significant differences between
the results.

The advantage of using the maximum rule is that, 
working in terms of logarithmic variables makes it possible to treat extremely small effective couplings 
occurring in a near-singular distribution. The maximum rule is also in the general spirit of the SDRG 
approach, where the flow is toward a singular distribution of couplings and the sum of contributions 
generated in the decimation steps becomes increasingly dominated by the maximum contribution as the RG flows 
toward the ground state.

%%%%%%%%%%%%%%%%%%%%%%%%%%%%%%% FIGS15  %%%%%%%%%%%%%%%%%%%%%%%%%%%%%%%%%
\begin{figure}[t]
\includegraphics[width=8cm, clip]{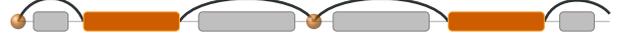}
\caption{\label{fig:danglingspins}
(Color online) A typical case where unpaired spins occur. 
The chain is periodic with two ends (in the figure) coupled to each other. 
The orange (gray) rectangular regions indicate blocks of $Q$-terms that are active (decimated). 
The black curves are effective $J$-couplings. When the active $Q$-terms are to be decimated, 
no effective coupling will be generated according to the decimation rule; 
thus the two spins (indicated by circles) are left unpaired.}
\end{figure}

%%%%%%%%%%%%%%%%%%%%%%%%%%%%%%%%%%%%%%%%%%%%%%%%%%%%%%%%%%%%%%%%%%%%%%

\subsection{Unpaired spins}
\label{sec:unpaired}

We have noticed that there is a small fraction of unpaired spins in the
approximate ground states of long random $Q$-chains. The cause for those
unpaired spins is the zero effective couplings in some cases of the $Q_3$-decimation; for
example, when a $Q_3$-term is decimated and the perturbative terms are solely a
pair of $J$-couplings [see Case (vi) in the $Q_3$-decimation procedure].
Fig.~\ref{fig:danglingspins} shows an example of such unpaired spins.  Since the
ground state of the $Q$-chain must be a spin-zero state, the rare
unpaired spins are certainly in a singlet state with a weak bond which is
ignored in the perturbative RG procedure. In short chains ($N\lesssim 200$),
we do not observe such unpaired spins. For longer chains where they do appear,
one reasonable way to account for them is simply to pair them up into singlets,
to ensure that the ground state on which we compute correlation functions is a singlet.
We have detected no significant differences between correlation functions computed
with these singlet pairings and with the unpaired spins left in the system, which
demonstrates that they do not play any significant role in practice.

%%%%%%%%%%%%%%%%%%%%%%%%%%%%%%%%%%%%%%%%%%%%%%%%%%%%%%%%%%%%%%%%%%%%%%%

\begin{table}[t]
\vskip4mm
\begin{center}
\begin{tabular}{ |r|l|l| } 
\hline
$N$~ & $C(N/2)$, this work & $C(N/2)$, Ref.~\onlinecite{laflorencie04} \\ 
\hline
8~    &   ~0.0918(2)  & ~0.0909(13) \\
12~   &   ~0.0534(1)  & ~0.0529(10) \\
16~   &   ~0.03540(6) & ~0.0346(7)  \\
20~   &   ~0.02533(3) & ~ \\
24~   &   ~0.01900(2) & ~0.0186(5) \\
28~   &   ~0.01479(2) & ~ \\
32~   &   ~0.01183(2) & ~0.0116(3) \\
36~   &   ~0.00970(1) & ~ \\
40~   &   ~0.00810(1) & ~ \\
44~   &   ~0.00685(1) & ~ \\
48~   &   ~0.00590(1) & ~0.00507(18) \\
52~   &   ~0.00511(1) & ~ \\
56~   &   ~0.00448(1) & ~ \\
60~   &   ~0.003960(7)~& ~ \\
64~   &   ~0.003521(3)~& ~ \\
72~   &   ~0.002843(3)~& ~\\
80~   &   ~0.00229(5)~ & ~ \\
100~  &   ~0.00152(4)~ & ~ \\
128~  &   ~0.00097(2)~ & ~ \\
~144~  &   ~0.00078(2)~ & ~\\
\hline
\end{tabular}
\end{center}
\vskip-3mm
\caption{Numerical values of the spin correlation function $C(r)=(-1)^r\langle {\bf S}_i {\bf S}_{i+r}\rangle$ at $r=N/2$
for the random-$J$ model with $d=1$, averaged over the reference location $i$ and disorder realizations. The results in this work were 
obtained using ground-state projector QMC calculations, while those from Ref.~\onlinecite{laflorencie04} were computed using finite-temperature 
QMC calculations at low temperatures (and adjusted by a factor $3/2$ to account for different definitions). The numbers in parentheses 
indicate the statistical error (one standard deviation) of the preceding digit.}
\label{tab1}
\end{table}

\section{Tabulated spin correlations}
\label{app:comp}
 
The disorder-averaged spin correlations were previously calculated using finite-temperature
QMC simulations at low temperatures in Ref.~\onlinecite{laflorencie04}. Results corresponding to our
$d=1$ distribution were shown in Fig.~8 (the $W=1$ data set) of Ref.~\onlinecite{laflorencie04}. 
To account for different prefactors
in definitions, the results there should be multiplied by $3/2$ to match our results in
Fig.~\ref{figj}. At first sight, the results in Ref.~\onlinecite{laflorencie04} appear to match well
the expected asymptotic $r^{-2}$ form without the log correction we have argued for
and which is required to fit our data in Fig.~\ref{figj}. However, by comparing with
our projector QMC results, we find a significant disagreement for the largest system size
($N=48$), which very likely is due to remaining finite-temperature effects in the
previous calculation. The $r^{-2}$ behavior does not match well the data if the
correct result for the largest system is used in Fig.~8 of Ref.~\onlinecite{laflorencie04}. 
The main conclusion in Ref.~\onlinecite{laflorencie04} regarding cross-over 
scaling with coupling distributions not extending to $J_i=0$ is not affected by this issue.

We list the results from Fig.~8 of Ref.~\onlinecite{laflorencie04} 
alongside our projector QMC results in Table \ref{tab1}. Good
agreement within statistical errors can be seen for all sizes smaller than $N=48$ 
(note, however, that all other results from \cite{laflorencie04} are also slightly below 
our current values, even though the deviations are within the error bars). 
Our error bars are also significantly reduced relative to those in Ref.~\onlinecite{laflorencie04}
and we have extended the range of reliably convergence considerably, up to $N=144$ for $d=1$.
We include these results for the benefit of future comparisons with other calculations.

\null

\end{document}